\documentclass[a4paper,twocolumn,11pt,unpublished]{quantumarticle}
\pdfoutput=1
\usepackage[utf8]{inputenc}
\usepackage[english]{babel}
\usepackage[T1]{fontenc}
\usepackage{amsmath}
\usepackage{hyperref}

\usepackage{booktabs}
\usepackage{amsmath,amssymb}

\usepackage[numbers,sort&compress]{natbib}
\usepackage{tikz}
\usepackage{lipsum}

\usepackage{listings}

\usepackage{amsmath}
\usepackage{amssymb}
\usepackage{hyperref}
\usepackage[resetlabels]{multibib}

\begin{document}

\title{Effects of coherent and incoherent measurement imperfections on multipartite quantum nonlocality and quantum key distribution}

\author{Qiong Wang}
\orcid{0009-0007-7492-8118}

\thanks{Qiong Wang and Wen-Long Qiao contributed equally to this work.}

\author{Wen-Long Qiao}
\orcid{0009-0008-9318-1037}

\author{Qing Chen}
\orcid{0000-0002-2191-6558}
\email{chenqing@ynu.edu.cn}

\author{Liu-Jun Wang}
\orcid{0000-0002-0886-2831}
\email{ljwangq@ynu.edu.cn}
\affiliation{
School of Physics and Astronomy and Yunnan Key Laboratory for
Quantum Information, Yunnan University, Kunming 650500, China}

\maketitle

\begin{abstract}
Multipartite Bell nonlocality is a central resource for device-independent quantum information protocols, but its practical certification is inevitably affected by imperfect measurements. 
We analyze how coherent angular misalignment and incoherent outcome flipping affect Bell-value degradation and nonlocality thresholds in $n$-partite GHZ states based on the Mermin, Svetlichny, and Mermin--Ardehali--Belinskii--Klyshko (MABK) inequalities.
Coherent misalignment produces periodic angular violation windows whose individual widths shrink with the number of parties.
In contrast, incoherent outcome flipping yields a single critical outcome-flipping probability, which increases with $n$ for MABK and the odd-$n$ Mermin inequalities, but decreases with $n$ for the Svetlichny inequality.
Connecting the degraded Bell values to asymptotic
Devetak--Winter key-rate bounds under a convex-combination attack model shows that secret-key generation imposes stricter constraints on measurement imperfections than nonlocality certification.
These results provide quantitative benchmarks for robust multipartite nonlocality certification and key-rate estimation under measurement imperfections.
\end{abstract}


\section{\label{sec:intro}Introduction}

Bell nonlocality excludes local hidden-variable descriptions of quantum correlations and represents a fundamental resource for device-independent quantum information processing, including randomness certification, quantum key distribution, and secret sharing \cite{Bell1964,RevModPhys.38.447,nonlocality-review2014,Pirandola2020,PhysRevLett.113.140501,PhysRevLett.128.110506,PhysRevA.111.012603,PhysRevA.101.052339,PhysRevA.100.012319,Pironio2010Nature}.
The Mermin and Mermin--Ardehali--Belinskii--Klyshko (MABK) inequalities provide representative criteria for certifying standard multipartite Bell nonlocality, whereas the Svetlichny inequality is used to certify genuine multipartite nonlocality beyond hybrid local-nonlocal models \cite{Mermin1990,Svetlichny1987,Ardehali1992,Belinskii1993,PhysRevLett.89.060401,PhysRevLett.88.170405}. These inequalities are naturally investigated in Greenberger-Horne-Zeilinger (GHZ) states, which generate full $n$-partite correlations and exhibit strong multipartite Bell violations \cite{Greenberger1989,Greenberger1990,Pan2000}.
Since Bell-inequality violations are also closely related to security analysis and key-rate bounds in device-independent protocols, understanding the robustness of multipartite Bell nonlocality in GHZ-state scenarios is important for both foundational studies and practical quantum information applications \cite{Acin2006,PhysRevLett.97.120405,Acin2007,PhysRevLett.127.050503,PRXQuantum}.

Practical multipartite Bell tests are inevitably affected by nonideal experimental conditions. Imperfect state preparation and state visibility can reduce the strength of the observed correlations \cite{Werner1989,PhysRevA.92.032305,PhysRevA.93.042308,PhysRevApplied.19.064003}, while channel loss further limits practical Bell and device-independent implementations \cite{PhysRevA.90.052314,PhysRevA.47.R747,Christensen2013,Xu2020}. Measurement-side imperfections such as finite detection efficiency, imperfect detector-response calibration, readout errors, and crosstalk have therefore motivated extensive studies of loophole-free Bell tests, detector tomography, readout-error mitigation, and crosstalk characterization \cite{2015Giustina,2015Hensen,2009Lundeen,Maciejewski2020,Bravyi2021,2021Nation,Sarovar2020}. In this work, we focus on measurement imperfections and consider two representative but physically distinct models: coherent angular misalignment of the implemented measurement bases and incoherent outcome flipping of the recorded results. The former is characterized by an angular deviation $\theta$, which rotates the implemented measurement bases at each party and modifies the phase dependence of multipartite full correlations. The latter is characterized by the correct-readout probability $p$.

These two types of measurement imperfections have been investigated in related but largely separate contexts. For coherent setting errors, Rosset et al. \cite{PhysRevA.86.062325} showed that imperfect alignment of measurement bases can compromise quantum-state tomography and entanglement-witness certification, while Cao et al.~\cite{Cao2024} extended this issue to genuine multipartite entanglement detection with imperfect measurements. Related studies on full-domain Bell correlations and sequential steering have further shown that quantum certification can depend sensitively on measurement settings and measurement sharpness \cite{PhysRevApplied.19.034049,PhysRevA.103.022421,Svegborn2025,Zhang2026PRL,Zhao2026PRA}. For incoherent measurement errors, Len et al. \cite{Len2022NC} analyzed quantum-metrology protocols with noisy detection and showed that measurement imperfections can constrain the attainable precision. In a setting closer to multipartite device-independent key generation, Xiang developed a Svetlichny-inequality-based multipartite cryptographic protocol~\cite{Xiang2023S} and further showed that imperfect measurement accuracy reduces both Bell violation and the extractable secret-key rate \cite{Xiang2023}. Despite these developments, a systematic comparison between coherent angular misalignment and incoherent outcome flipping in $n$-partite GHZ Bell tests remains lacking. In particular, it remains unclear how these two physically distinct mechanisms affect the Mermin, MABK, and Svetlichny inequalities differently, and how the resulting degraded Bell values influence key-rate bounds based on a convex-combination attack model.

Here we make a systematic analytical comparison of coherent angular misalignment and incoherent outcome flipping in $n$-partite GHZ Bell-test scenarios. For the Mermin and MABK inequalities, which certify standard multipartite Bell nonlocality, and the Svetlichny inequality, which certifies genuine multipartite nonlocality, we derive the degraded Bell values and the corresponding critical thresholds under both imperfection models.
For coherent misalignment, we determine the angular parameter ranges over which the degraded Bell values remain above the relevant classical bounds and show that these ranges appear as periodic violation windows whose widths depend on both the number of parties and the Bell inequality considered.
For incoherent outcome flipping, we obtain a single critical threshold in the flipping probability, reflecting the stochastic attenuation of multipartite full correlations. We further connect the degraded Svetlichny Bell values to asymptotic Devetak--Winter key-rate bounds under a convex-combination attack model. This Svetlichny-based key-rate analysis shows that positive secret-key generation imposes stricter requirements on angular alignment and correct-readout probability than the certification of genuine multipartite nonlocality alone.

The remainder of this paper is organized as follows. In Sec.~\ref{paragraph2}, we introduce the ideal GHZ-state Bell scenario and the two measurement-imperfection models. In Sec.~\ref{paragraph3}, we derive the degradation of Mermin, Svetlichny, and MABK violations under coherent angular misalignment and incoherent outcome flipping, and obtain the corresponding nonlocality certification thresholds. In Sec.~\ref{paragraph4}, we relate the degraded Bell values to asymptotic key-rate bounds and discuss the role of Werner-state visibility. Finally, Sec.~\ref{paragraph5} summarizes the main conclusions and discusses possible extensions.

\section{\label{paragraph2}Multipartite Bell Inequalities and Measurement-Imperfection Models}

In idealized quantum information protocols, measurements are traditionally modeled as perfect projective operators. However, practical implementations are invariably subject to hardware imperfections, including basis misalignments, environmental noise, and finite instrument precision.
Although different measurement imperfections may appear as similar attenuations of the measured signal at the ensemble level, their physical mechanisms and actions on the Bell operator are distinct.
To analytically assess these impacts, we consider two representative  measurement-imperfection models: coherent angular misalignment and incoherent outcome flipping. The former behaves as a unitary rotation in the operator space, while the latter represents a classical stochastic process described by the positive-operator-valued-measure (POVM) framework.

\subsection{\label{2.A}Multipartite GHZ states and associated Bell-type inequalities}

Before introducing measurement imperfections, we briefly summarize the ideal multipartite GHZ Bell-test scenario. 
Multipartite entanglement and its certification have been extensively studied in both entanglement-witness and Bell-inequality frameworks \cite{RevModPhys.81.865,Guhne2009,PhysRevLett.88.230406}.
To express the Bell operators compactly in terms of complex conjugate combinations, and without loss of generality up to local unitary transformations, we consider the $n$-partite GHZ state with a relative $\pi/2$ phase:
\begin{equation}
    \left |\text{GHZ}_{n} \right \rangle = \frac{\left |0 \right \rangle^{\otimes n} + i\left|1 \right\rangle^{\otimes n}}{\sqrt{2}},
    \label{eq:1}
\end{equation}
which is locally unitarily equivalent to the standard real-valued GHZ state. 
The local dichotomic measurements are chosen in the $x-y$ plane of the Bloch sphere, yielding a phase-sensitive full $n$-partite correlation that underpins the maximal violation of multipartite Bell inequalities.

In this work, we focus on three families of full-correlation Bell inequalities: the Mermin inequality, the Svetlichny inequality, and the MABK inequality. They are summarized as follows.

\subsubsection{Mermin inequality}

For the Mermin inequality \cite{Mermin1990}, the ideal local observables are chosen as $A_0^{(j)} = \sigma_x, \quad A_1^{(j)} = \sigma_y$. The Mermin Bell operator used in this work can be written as

\begin{equation}
    M_n = \frac{1}{2i} \left( \prod_{j=1}^{n} \zeta^{+} - \prod_{j=1}^{n} \zeta^{-} \right),
\label{eq:2}
\end{equation}
where $\zeta^{\pm} = A_0^{(j)} \pm i A_1^{(j)}$. 
Under the ideal GHZ measurement configuration, the maximal quantum value is $S_Q^{\text{Mer}} = 2^{n-1}$. The corresponding local-hidden-variable bound is parity dependent:
\begin{equation}
    S_{C}^{\text{Mer}} = 
    \begin{cases}
        2^{\left (n-1\right )/2} & n \text{ odd}, \\
        2^{n/2} & n \text{ even}.
    \end{cases}
    \label{eq:3}
\end{equation}

This parity dependence is important because it makes the standard Mermin inequality suboptimal for even $n$, which motivates the use of the MABK inequality in the even-party case.

\subsubsection{Svetlichny inequality}

For the $n$-partite Svetlichny inequality \cite{Svetlichny1987,PhysRevLett.89.060401}, the following expression always holds:

\begin{equation}
    \left|\left\langle S^{\pm}_n \right\rangle\right| =\left|\sum_I v^{\pm}_{t(I)}E(x_1,x_2 \dots x_n)\right| \leq 2^{n-1},
    \label{eq:4}
\end{equation}
where $I=(x_1,x_2,\ldots,x_n)$ denotes the input string,
with $x_i\in\{0,1\}$ labeling the measurement setting chosen
by the $i$th party. The quantity
$t(I)=\sum_{i=1}^{n}x_i$ counts the number of parties choosing
their second measurement setting, $x_i=1$, and the corresponding
coefficient is
$v_{t(I)}^{\pm}=(-1)^{t(I)[t(I)\pm1]/2}$. The $n$-partite Svetlichny operator $S^{-}_n$ can be expressed as the sum of $M_n$ and $N_n$  \cite{PhysRevLett.88.230406}. 
For a single party, the operators are initialized as $M_1 = A_0^{(1)}$ and $ N_1 = A_1^{(1)}$ . For $k \ge 2$, the recursion relations are $ M_k = M_{k-1} A_0^{(k)} - N_{k-1} A_1^{(k)}$ and $ N_k = M_{k-1} A_1^{(k)} + N_{k-1} A_0^{(k)}$. Equivalently, they can be expressed as
\begin{subequations}
    \label{eq:5}
    \begin{equation}
        N_n=\frac{1}{2}\left ( \prod_{j=1}^{n} \zeta^{+}  + \prod_{j=1}^{n} \zeta^{-} \right ) ,
    \end{equation}
    \label{5a}
    \begin{equation}
        M_n=\frac{1}{2i}\left ( \prod_{j=1}^{n} \zeta^{+}  - \prod_{j=1}^{n} \zeta^{-} \right )  .
    \label{5b}
    \end{equation}
\end{subequations}

The maximal violation is obtained when one designated party, say party $m$, uses the rotated observables $A_0^{(m)} = \frac{1}{\sqrt{2}} (\sigma_y - \sigma_x), \quad A_1^{(m)} = \frac{1}{\sqrt{2}} (\sigma_x + \sigma_y)$, while all other parties use $A_0^{(j)} = \sigma_x, \quad A_1^{(j)} = \sigma_y \quad (j \neq m)$. 
With this convention, the Svetlichny quantum maximum is $ S_Q^{\text{Sve}} = 2^{n-1/2}$. Equivalently, the ratio of the quantum bound to the classical maximum is $S_Q^{\text{Sve}}/S_C^{\text{Sve}}=\sqrt{2}$. 
Unlike the standard Mermin inequality, the Svetlichny inequality is used to certify genuine multipartite nonlocality rather than merely standard multipartite Bell nonlocality.

\subsubsection{MABK inequality}
The MABK inequality provides an optimized multipartite Bell criterion that removes the even-$n$ suboptimality of the standard Mermin inequality \cite{Mermin1990,Ardehali1992, Belinskii1993}. 
The MABK Bell operator is most conveniently constructed through the Belinskii-Klyshko recurrence relations. 
For a single party ($n=1$), the operators are initialized as $B_1 = A_0^{(1)}$ and $\quad B_1' = A_1^{(1)}$.
For any $k \ge 2$, the operators are generated recursively by
\begin{equation}
    B_k = \frac{1}{2} [ B_{k-1} ( A_0^{(k)} + A_1^{(k)}) + B_{k-1}'( A_0^{(k)} - A_1^{(k)} ) ].
\label{eq:6}
\end{equation}

We use the normalized MABK convention, in which a common overall normalization factor is applied to all correlation terms of the MABK Bell parameter, yielding the local-hidden-variable bound $S_C^{\text{MK}} = 1$. 
The ideal quantum maximum is $ S_Q^{\text{MK}} = 2^{(n-1)/2}$. It should be noted that this normalization factor is only a matter of convention and does not alter the physical content of the MABK inequality. If the overall fractional prefactor is omitted, the Bell parameter is globally rescaled, and consequently both the classical and quantum bounds are rescaled by the same factor, as summarized in Table \ref{tab:table1}.

For odd $n$, the MABK inequality coincides with the optimized form of the Mermin inequality. For even $n$, it provides a stronger criterion than the standard Mermin inequality and is therefore more suitable for comparing measurement robustness across different system sizes.

In this work, we denote by $S$ the unnormalized Bell value, by $S_C$ the corresponding local-hidden-variable bound, and by $S_Q$ the maximal quantum value. 
For comparing different inequalities, we also introduce the normalized violation ratio $\bar S=S/S_C$, for which the classical bound is always $\bar S_C=1$. Unless otherwise specified, the analytical derivations below are written in terms of the unnormalized quantities $S$, $S_C$, and $S_Q$.

\begin{table}[htp]
\centering
\caption{\label{tab:table1}
Classical bounds and maximal quantum values of the multipartite
Bell inequalities considered in this work. Here $S$ denotes the
unnormalized Bell value, while $\bar S=S/S_C$ denotes the
normalized violation ratio. The normalized quantum value is
$\bar S_Q=S_Q/S_C$.
}
\begin{tabular*}{\columnwidth}{@{}c@{\extracolsep{\fill}}cccc@{}}
\toprule
Inequality & \(S_C\) & \(S_Q\) & \(\bar S_Q=\frac{S_Q}{S_C}\) \\
\midrule
Mermin, $n$ odd & \(2^{\frac{(n-1)}{2}}\) & \(2^{n-1}\) & \(2^{\frac{(n-1)}{2}}\) \\[3pt]
Mermin, $n$ even & \(2^{n/2}\) & \(2^{n-1}\) & \(2^{\frac{(n-2)}{2}}\) \\[3pt]
Svetlichny & \(2^{n-1}\) & \(2^{n-\frac{1}{2}}\) & \(\sqrt{2}\) \\[3pt]
MABK, $n$ odd & \(2^{\frac{(n-1)}{2}}\) & \(2^{n-1}\) & \(2^{\frac{(n-1)}{2}}\) \\[3pt]
MABK, $n$ even & \(2^{\frac{n}{2}}\) & \(2^{n-\frac{1}{2}}\) & \(2^{\frac{(n-1)}{2}}\)\\
\bottomrule
\end{tabular*}
\end{table}

These ideal Bell values provide the reference points for evaluating the effects of measurement imperfections. In the next two subsections, we introduce two physically distinct imperfection models: coherent angular misalignment and incoherent outcome flipping.

\subsection{\label{2.B}Models of imperfect measurements}

In realistic multipartite Bell experiments, the implemented local measurements are not identical to the ideal projective observables assumed in the theoretical Bell-test model. In this work, we consider two representative and experimentally relevant measurement imperfections: coherent angular misalignment and incoherent outcome flipping. The former corresponds to a systematic angular deviation of the measurement basis, whereas the latter describes a classical error in the recorded measurement outcome.

\textit{Coherent angular misalignment.---}
Measurement misalignment typically arises from geometric tilts in the apparatus or imperfect calibration of the local measurement bases. Physically, this represents a coherent systematic error. In the Heisenberg picture, this imperfection is represented by a unitary rotation of the measurement observable within the operator subspace spanned by the intended observable and its orthogonal component.

Specifically, let $\theta$ parameterize the degree of angular deflection from the optimal measurement direction. Let $A_i$ denote the intended ideal measurement operator; under a coherent deflection, the actual observable evolves into

\begin{equation}
    \hat{A}_i' = \cos(\theta)A_{i} + \sin(\theta)A_{i}^\perp ,
    \label{eq:7}
\end{equation}
where $A_{i}^\perp$ represents the operator component orthogonal to the intended basis. 

In this model, the measurement remains projective and the observable eigenvalues are unchanged.
Instead of simply attenuating the signal, this misalignment induces a phase-dependent modulation. In multipartite systems, such as $n$-partite GHZ states, this coherent superposition triggers complex interference effects, leading to a non-trivial parity-dependent robustness. Such errors primarily require hardware-level calibration.

\textit{Incoherent outcome flipping.---} In contrast to the geometric nature of misalignment, outcome-flipping models an incoherent stochastic process. This captures classical statistical errors arising from non-ideal detector performance, such as dark counts, readout noise, or qubit relaxation during the measurement process. 

Mathematically, this stochastic readout process can be described within the POVM framework. We denote by $p$ the correct-readout probability, namely the probability that the intended outcome $\pm1$ is recorded correctly, while $1-p$ is the outcome-flipping probability, corresponding to an erroneous inversion of the recorded outcome. For an ideal projective observable $A_i=|a\rangle\langle a|-|\bar a\rangle\langle \bar a|$, the effective POVM elements associated with the two detector clicks are given by
\begin{equation}
    \begin{cases}
        E_a = p |a\rangle \langle a| + (1 - p)|\bar{a}\rangle \langle\bar{a}| \\ 
        E_{\bar{a}} = p |\bar{a}\rangle \langle\bar{a}| + (1 - p)|a\rangle \langle a| 
    \end{cases}.
    \label{eq:8}
\end{equation}
Experimentally, this process can be regarded as a classical readout confusion channel. Provided that the correct-readout probability $p$ is independently calibrated, the ideal expectation values can in principle be recovered by applying the inverse transition matrix, as in readout-error mitigation.

\section{\label{paragraph3}Multipartite Bell Violations under Imperfect Measurements}

We now analyze how the two measurement-imperfection models modify multipartite
Bell violations in $n$-partite GHZ states. In the ideal case, the chosen measurement settings maximize the corresponding Bell values. Under imperfect measurements, these values are degraded and may fall below the relevant nonlocality thresholds.

\subsection{\label{3.A}Coherent angular misalignment}

Using the Bell operators defined in Sec.~\ref{2.A}, we now evaluate how coherent angular misalignment modifies the Mermin, Svetlichny, and MABK Bell values. The local observables are replaced by the imperfect operators introduced in Sec.~\ref{2.B}.
In the following, we discuss the scenario where all parties experience identical errors, while the case of party-dependent errors is addressed in Appendix \ref{Appendix:A}. 
For the former, we consider the deflection angles to be $\theta_1$ for the first measurement and $\theta_2$ for the second measurement.

For the Mermin inequality, the parties measure two local observables, $\sigma_x$ and $\sigma_y$. Under coherent angular misalignment, the implemented measurement operators evolve into

\begin{subequations}
    \label{eq:9}
    \begin{equation}
        \tilde{\sigma_{x}}  =\cos{\theta _{1}} \sigma_{x}+\sin{\theta _{1}}  \sigma_{y},
    \end{equation}
    \label{9a}
    \begin{equation}
        \tilde{\sigma_{y}}  =\cos{\theta _{2}} \sigma_{y}-\sin{\theta _{2}}  \sigma_{x}.
    \label{9b}
    \end{equation}
\end{subequations}
In the context of $n$-partite Mermin inequalities, the condition $\theta=0$ characterizes an ideal measurement configuration free of misalignment. Eq. (\ref{eq:9}) provides an explicit realization of the rotation model in Eq. (\ref{eq:7}) for the two ideal observables $\sigma_x$ and $\sigma_y$. A positive $\theta_1$ rotates $\sigma_x$ counterclockwise in the $x$-$y$ plane. Since $\tilde{\sigma}_y=\cos\theta_2\sigma_y-\sin\theta_2\sigma_x$ corresponds to an azimuthal angle $\pi/2+\theta_2$, a positive $\theta_2$ also rotates $\sigma_y$ counterclockwise. In particular, when $\theta_1=\theta_2$, the two implemented observables remain orthogonal, corresponding to an orthogonality-preserving common rotation.
\begin{figure*}[t!]
    \centering
    \includegraphics[width=\textwidth]{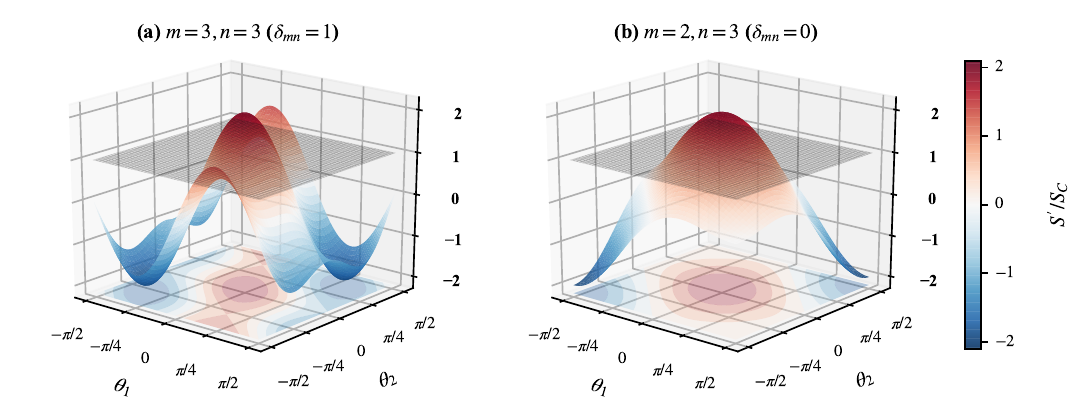} 
    \caption{
    Evolution of the normalized Mermin violation $S'/S_C$ for an $n=3$ GHZ state as a function of the local misalignment angles $\theta_1$ and $\theta_2$. Panels (a) and (b) illustrate the full-system ($m=3$) and partial-system ($m=2$) misalignment scenarios, respectively. The horizontal gray mesh plane indicates the classical local-realistic limit at $1$. The diagonal interference patterns in the violation landscape arise from the underlying dependence on the co-planar and differential deflection parameters $\beta$ and $\alpha$, highlighting the distinct sensitivity of the Mermin inequality to symmetric versus asymmetric measurement errors.}
    \label{fig:Mermin_Deflection}
\end{figure*}

To streamline subsequent calculations, we introduce the complex combinations $\ell_{\pm}=\tilde{\sigma}_{x}\pm i\tilde{\sigma}_{y}$, as well as the auxiliary angles $\alpha=(\theta_1-\theta_2)/2$ and $\beta=(\theta_1+\theta_2)/2$. We then analyze how the expectation value of the Bell inequality changes when $m$ out of the $n$ measurement parties are affected by imperfect measurements, with $m\leq n$. For the case of coherent angular misalignment, substituting the imperfect observables in Eq. (\ref{eq:9}) into $\ell_{\pm}$ and evaluating the resulting full $n$-partite correlations on the GHZ state gives

\begin{equation}
\begin{split}
    \left\langle \prod_{j=1}^{n}\ell_+^{(j)} \right\rangle =& i 2^{n-1} [ \left(\cos\alpha e^{         -i\beta}\right)^{m}  \\
    &-\delta_{m,n} \left(i \sin\alpha e^{i\beta}\right)^{n} ].
\end{split} 
\label{eq:10}
\end{equation}
Note that $\langle \prod_{j=1}^{n}\ell_-^{(j)} \rangle$ is simply the complex conjugate of $\langle \prod_{j=1}^{n}\ell_+^{(j)}\rangle$, as $\langle \prod_{j=1}^{n}\ell_-^{(j)} \rangle = \langle \prod_{j=1}^{n}\ell_+^{(j)}\rangle^\ast$. Here $\delta_{m,n}$ is the Kronecker delta. It indicates that the second term contributes only when all $n$ parties are faulty. For partial faulty-node configurations with $m<n$, this global interference term is absent. In the presence of measurement deflection, the expectation value of the Mermin operator $S'$ is determined by the number of faulty nodes $m$ ($m \leq n$). Utilizing the derived evolution of the measurement bases, the expectation value for $n$-partite GHZ states is given by

\begin{equation}
\begin{split}
    S'_{\rm Mer} =& 2^{n-1} [ \cos^m(\alpha ) \cos(m\beta ) \\    
    & - \delta_{m,n} \sin^n(\alpha ) \cos(n\beta + \frac{n\pi}{2}) ].
\end{split}
\label{eq:11}
\end{equation}
For $m<n$, the Kronecker-delta term in Eq. (\ref{eq:11}) vanishes, so that
$S_{\rm Mer}'=2^{n-1}\cos^m(\alpha)\cos(m\beta).$ Thus, partial faulty-node configurations exhibit a smooth modulation controlled by the number of imperfect measurements.
In the fully faulty case $m=n$, the additional interference term survives and gives rise
to the parity-dependent structure shown in Fig. \ref{fig:Mermin_Deflection}.

Incorporating the measurement imperfections parameterized by Eq. (\ref{eq:7}), we obtain the evolution of the Svetlichny violation $S'$ as a function of the misalignment angle

\begin{equation} 
\begin{split}
    S'_{\rm Sve} = &2^{n-\frac{1}{2}} [ \cos^m(\alpha) \cos(m\beta)  \\
    &+\delta_{m,n} \sin^n(\alpha) \cos(n\beta + \frac{(n-1)\pi}{2}) ].
\end{split}
\label{eq:12}
\end{equation}

Although Fig.~\ref{fig:Mermin_Deflection} specifically illustrates the Mermin inequality, the same coherent-interference mechanism also appears in other symmetrically structured full-correlation criteria, such as the Svetlichny inequality. Eq. (\ref{eq:11}) and~(\ref{eq:12}) show that the Mermin and Svetlichny violations have the same leading factor $\cos^m(\alpha)\cos(m\beta)$, but differ in the global interference term present for $m=n$. This term has different relative signs and parity-dependent phases, $n\pi/2$ for Mermin and $(n-1)\pi/2$ for Svetlichny, producing different violation landscapes under coherent misalignment.
Thus, the degradation is governed not only by the
error magnitude, but also by the algebraic structure of the Bell operator and the symmetry of the faulty-node configuration.
The MABK inequality provides an optimized full-correlation Bell criterion that coincides with the Mermin criterion for odd $n$ and removes the even-$n$ suboptimality of the standard Mermin inequality.
For the MABK inequality, the classical bound is normalized to unity, while the maximum quantum expectation value reaches $2^{(n-1)/2}$ . Incorporating the measurement imperfections, the degraded unnormalized MABK violation can be expressed as

\begin{equation}
\begin{split}
    S'_{\rm MK}= &S_Q[\cos^m(\alpha)\cos(m\beta)\\
    &+(-1)^{n}\delta_{m,n}\sin^n(\alpha)\\
    &\times\cos(n\beta+\frac{(n+1-2^{\Lambda })\pi}{2})],
\end{split}
\label{eq:13}
\end{equation}
where $\Lambda=(n+1) \mod 2$. 

In the threshold analysis below, unless stated otherwise, we restrict attention to the orthogonality-preserving rotation model, $\theta_1=\theta_2\equiv\theta$, for which the two effective angular parameters reduce to $\alpha=0$ and $\beta=\theta$. Under this rotation model, the degraded expectation value for all the above inequalities can be written in the compact form $S'=S_Q\cos(m\theta)$, where $m\le n$ denotes the number of faulty measurement nodes.
By equating the degraded Bell value to the classical bound $S_C$, we obtain the critical misalignment angle $\theta_{\rm cr}$ for preserving a Bell violation. Throughout this threshold analysis, we restrict our attention to the positive-violation branch, where the degraded Bell value satisfies $S'=S_Q\cos(m\theta)>S_C$. The opposite branch, corresponding to $S'<-S_C$, is not considered in the present convention. Specifically, for a configuration with $m$ imperfect parties, the critical angle is $\theta_{\rm cr} = \frac{\pm\arccos(S_C/S_Q)+2k\pi}{m}$. In the fully imperfect scenario where all parties suffer from this orthogonal deflection ($m=n$), the solution strictly yields $\theta_{\rm cr} = \frac{\pm\arccos(S_C/S_Q)+2k\pi}{n}$, where $k \in \mathbb{Z}$ is an arbitrary integer.

By substituting the respective classical and quantum bounds ($S_C$ and $S_Q$) summarized in Table \ref{tab:table1}, we obtain the explicit general solutions for the critical misalignment angles associated with each inequality. 

For the Mermin inequality, the critical thresholds exhibit a distinct parity dependence, summarized as
\begin{equation}
    \theta_{\rm cr}^{\text{Mer}} = 
    \begin{cases} 
        \dfrac{\pm\arccos\left(2^{(1-n)/2}\right) + 2k\pi}{n} & n \text{ is odd} \\[12pt]
        \dfrac{\pm\arccos\left(2^{(2-n)/2}\right) + 2k\pi}{n} & n \text{ is even}
    \end{cases}.
    \label{eq:14}
\end{equation}

In contrast, the MABK inequality provides a unified threshold applicable to any $n$:
\begin{equation}
    \theta_{\rm cr}^{\text{MK}} = \frac{\pm\arccos\left(2^{(1-n)/2}\right) + 2k\pi}{n},
    \label{eq:15}
\end{equation}
which inherently coincides with the Mermin threshold specifically for odd $n$. 

Finally, the critical angle for the Svetlichny inequality exhibits a universal inverse scaling with the system size $n$, modulated by the periodicity parameter $k$
\begin{equation}
    \theta_{\rm cr}^{\text{Sve}} = \frac{\pm\arccos(1/\sqrt{2}) + 2k\pi}{n} = \frac{\pm \pi/4 + 2k\pi}{n}.
    \label{eq:16}
\end{equation}

\begin{figure}[t!]
    \centering
    \includegraphics[width=\columnwidth]{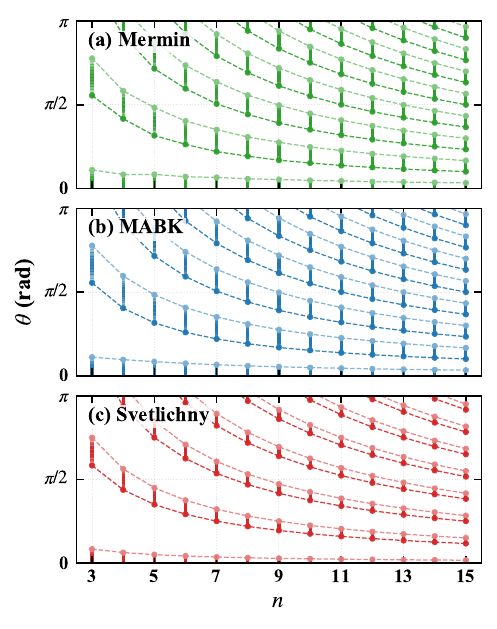} 
    \caption{
Full angular-domain violation regions under full-system ($m=n$) coherent angular misalignment for $n$-partite GHZ states, evaluated at $\beta=\theta$.
The solid vertical segments highlight the discrete angular intervals where the Bell inequality is violated ($S' > S_C$). 
The dark and light dashed lines represent the analytical lower and upper critical misalignment angles ($\theta_{\rm cr}$), marking the analytical phase boundaries for entering and leaving the nonlocal regimes, respectively. Blank regions between the segments denote regions where the corresponding Bell inequality is not violated.
As $n$ increases, the violation windows become narrower and denser, reflecting the $2\pi/n$ periodicity of the accumulated phase $n\theta$. Notably, the MABK and Mermin thresholds coincide for odd $n$, whereas the Svetlichny inequality exhibits a universal $1/n$ inverse scaling.}
    \label{fig:2}
\end{figure}

Fig.~\ref{fig:2} shows the violation regions satisfying $S'>S_C$ over the angular
domain $\theta\in[0,\pi]$ for the fully faulty case $m=n$. The resulting patterns reflect the dependence of robustness on the
algebraic structure of the Bell operators.
For the Mermin inequality [Fig.~\ref{fig:2}(a)], a parity-dependent zigzag pattern appears, with narrower violation bands for even $n$. This vulnerability stems from the suboptimal classical bound of the Mermin operator in even-party systems. The MABK criterion [Fig.~\ref{fig:2}(b)] resolves this defect by optimizing the algebraic recurrence, thereby yielding a smooth, monotonically decaying robustness envelope. 
It coincides with the Mermin inequality for odd $n$ while significantly broadening the tolerable parameter space for even $n$.

In contrast, Fig.~\ref{fig:2}(c) 
shows that the Svetlichny criterion is more sensitive to coherent misalignment.
Designed to rule out hybrid local--nonlocal correlation
models, the Svetlichny inequality is constrained by a fixed
hybrid-bound-to-quantum-maximum ratio of $1/\sqrt{2}$.
Consequently, its violation bands are substantially narrower than those of the standard nonlocality criteria. Furthermore, as the system size $n$ increases, the $2\pi/n$ periodicity of the interference term 
generates increasingly dense violation windows whose widths decrease with $n$.

In the fully faulty case, the violation windows satisfying $S'>S_C$ are centered at $ \theta_k=2k\pi/n$. The centers of the Mermin, MABK, and Svetlichny violation windows therefore coincide for the same integer $k$, while their widths are set by the corresponding ratio $S_C/S_Q$.

For a complete violation window, the angular length is $2\arccos(S_C/S_Q)/n$. In the large-$n$ limit, the half-widths of the Mermin and MABK windows approach $\pi/(2n)$, whereas the Svetlichny half-width is $\pi/(4n)$. Thus each complete Mermin or MABK violation segment is asymptotically twice as long as the corresponding Svetlichny segment.
Although each individual segment shrinks as $1/n$, the number of positive-violation windows in $[0,\pi]$ increases linearly with $n$. Consequently, the total positive-violation measure approaches $L_{\rm Mer/MK}\to \pi/2$ for the Mermin/MABK criteria and $L_{\rm Sve}\to \pi/4$ for the Svetlichny criterion, producing the narrower but denser pattern shown in Fig.~\ref{fig:2}. We note that these measures refer only to the positive-violation branch adopted in the present convention. If one instead considers the absolute-violation condition $|S'|>S_C$, which also includes the opposite branch $S'<-S_C$, the half-width of each window is unchanged, but the centers become $\theta_k=k\pi/n$. The total violation measure is then doubled, approaching $L_{\rm Mer/MK}\to \pi$ and $L_{\rm Sve}\to \pi/2$.

\subsection{Incoherent outcome flipping and correlation scaling}

We next consider incoherent outcome flipping. In this model, the measurement basis is assumed to be correctly implemented, while each recorded binary outcome is flipped with probability $1-p$, where $p$ denotes the correct-readout probability. Unlike coherent angular misalignment, this imperfection does not change the measurement direction or the phase structure of the local observable. Instead, it represents a stochastic readout error arising from nonideal detector performance, such as dark counts, readout noise, or relaxation during the measurement process.

Because the flipping process is independent of the chosen measurement basis, its effect on multipartite correlations is expected to be uniform. Specifically, every full $n$-partite correlation acquires the same multiplicative factor $(2p-1)^n$. To show this explicitly, we use the probabilistic formulation of the Mermin inequality. For the multipartite scenario, the Mermin inequality can be written as
\begin{equation}
\begin{split}
    &2\sum_{x_i}P\left(\sum_i^na_i=\sum_{i< j\leq n}x_ix_j|x_1,x_2 \dots x_n\right)-2^{n-1}\\
    &\leq S_C.
\end{split}
\label{eq:17}
\end{equation}
This study focuses on symmetric full-correlation probabilities, assuming that the conditional probability $P\left(\sum_{i=1}^n a_i = \sum_{i<j \leq n} x_i x_j | x_1, x_2 \dots x_n \right)$ has the same readout-error scaling for all relevant measurement settings $(x_1, x_2 \dots x_n)$. Not all inequalities can be written in this form; this representation is restricted to symmetric multipartite full-correlation inequalities. 
This situation is explained in Appendix \ref{Appendix:B}. The corresponding local-causality bound and the maximum quantum value of the Mermin inequality are given by

\begin{subequations}
\label{eq:18}
\begin{equation}
\begin{split}
    &\frac{1}{2^{n-1}} \sum_{\left \{  x_i \right \}  } P\left( \sum_{i}^n a_i = \sum_{i< j\leq n} x_i x_j \,\middle|\, x_1 \dots x_n \right) \\
    &\leq \frac{S_C}{2^{n}} + \frac{1}{2},
\end{split}
\label{eq:18a}
\end{equation}
\begin{equation}
\begin{split}
    &\frac{1}{2^{n-1}} \sum_{\left \{  x_i \right \}} P\left( \sum_{i}^{n} a_i = \sum_{i< j\leq n} x_i x_j \,\middle|\, x_1 \dots x_n \right) \\
    &\leq \frac{S_Q}{2^{n}} + \frac{1}{2}.
\end{split}
\label{eq:18b}
\end{equation}
\end{subequations}

Here, $x_i\in\{0,1\}$ labels the measurement setting chosen by the $i$-th party, with $x_i=0$ and $x_i=1$ corresponding to the two local bases used in the Mermin test. The set $\{x_i\}$ denotes input strings $(x_1,x_2,\dots,x_n)$ satisfying $\sum_i x_i \equiv 0 \pmod 2$.
The binary outcome $A_i=\pm1$ is equivalently represented by $a_i=(1-A_i)/2\in\{0,1\}$. The conditional probability $P \left( \sum_{i}^{n} a_i = \sum_{i< j\leq n} x_i x_j \,\middle|\, x_1, x_2, \dots, x_n \right)$ therefore denotes the probability that the observed output parity satisfies the Mermin parity constraint. Under perfect measurements on an $n$-partite GHZ source, appropriate measurement settings achieve the Mermin quantum value $S_Q=2^{n-1}$.

Under the outcome-flipping model introduced in Sec.~\ref{2.B}, let $P' \left( \sum_{i}^{n} a_i = \sum_{i< j\leq n} x_i x_j \,\middle|\, x_1, x_2 \dots x_n \right)$ denote the conditional probability distribution of the observed outcomes under imperfect measurements. Crucially, the correct-readout probability $p$ is assumed to be independent of the measurement settings chosen by the parties, meaning this local readout accuracy remains invariant across all basis choices.

Under both ideal and imperfect measurement scenarios, the measurement configuration is denoted by $(x_1 \dots x_n)$. Let $M_{\text{tot}}$ be the total number of events, such that $M_{\text{tot}} = M_1 + M_2 = M'_1 + M'_2$. Here, $M_1$ and $M'_1$ represent the number of events satisfying the condition $\sum_{i=1}^n a_i = \sum_{i< j\leq n} x_i x_j$, while $M_2$ and $M'_2$ correspond to those that do not. Consequently, the probability distributions for the ideal and imperfect cases are written as

\begin{subequations}
\label{eq:19}
\begin{equation}
\begin{split}
    \frac{M_{1}}{M_{tot}} = P\left(\sum_{i}^{n} a_i = \sum_{i< j\leq n} x_i x_j \,\middle|\, x_1 \dots x_n \right),
\end{split}
\label{eq:19a}
\end{equation}
\begin{equation}
\begin{split}
    \frac{M_{1}'}{M_{tot}'} = P'\left(\sum_{i}^{n} a_i = \sum_{i< j\leq n} x_i x_j \,\middle|\, x_1 \dots x_n \right).
\end{split}
\label{eq:19b}
\end{equation}
\end{subequations}

The relationship between the two probability distributions is then
\begin{widetext}
    \begin{equation}
    \begin{split}
    \frac{1}{2^{n-1}} \sum_{x_i} P'\left( \sum_{i}^{n} a_i = \sum_{i< j\leq n} x_i x_j \,\middle|\, x_1 \dots x_n \right) 
    &= (2p-1)^n \times \frac{1}{2^{n-1}} \sum_{x_i} P\left( \sum_{i=1}^{n} a_i = \sum_{i< j\leq n} x_i x_j \,\middle|\, x_1 \dots x_n \right) \\
    &+ \frac{1-(2p-1)^n}{2}.
    \label{eq:20}
    \end{split}
    \end{equation}
\end{widetext}

For the Mermin inequality, incoherent outcome flipping of the recorded outcomes
reduces the observed quantum value from $S_Q$ to $S'_Q$, and Eq. (\ref{eq:18b}) is modified as follows

\begin{equation}
\begin{split}
    &\frac{1}{2^{n-1}} \sum_{x_i}P'\left( \sum_{i}^{n} a_i = \sum_{i< j\leq n} x_i x_j \,\middle|\, x_1, x_2 \dots x_n \right) \\
    &\le \frac{S'_Q}{2^{n}} + \frac{1}{2}.
\end{split}
\label{eq:21}
\end{equation}

The dependence of the quantum violation on the correct-readout probability $p$ is thus given by

\begin{equation}
S'_Q = S_Q (2p-1)^n.
\label{eq:22}
\end{equation}

\begin{figure}[t!]
    \centering
    \includegraphics[width=\columnwidth]{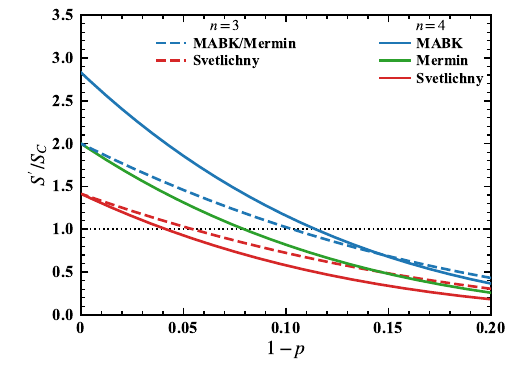} 
    \caption{
Normalized Bell violation $S'/S_C$ versus the outcome-flipping
probability $1-p$ for $n=3$ and $n=4$ GHZ states. Dashed and
solid curves correspond to $n=3$ and $n=4$, respectively. The
horizontal dotted line indicates the classical threshold
$S'/S_C=1$. The monotonic decay of all curves follows from the
uniform correlation scaling factor $(2p-1)^n$. For $n=3$, the
MABK and Mermin curves overlap because the two inequalities are
equivalent for odd $n$. In the even-party case $n=4$, the MABK
inequality retains a larger violation than the standard Mermin
inequality, while the Svetlichny inequality remains the most restrictive
criterion for genuine multipartite nonlocality.
    }
    \label{fig:3}
\end{figure}
 Fig.~\ref{fig:3} plots the normalized violation $S'/S_C$ versus the outcome-flipping probability $1-p$ for GHZ states with $n=3$ and $n=4$. All curves exhibit monotonic decay due to the uniform attenuation of multipartite correlations.
 
The critical correct-readout probability required to violate the inequality is therefore
\begin{equation}
    p_{\rm cr}=\frac{\left( S_{C} / S_{Q} \right)^{1/n} + 1}{2}.
\label{eq:23}
\end{equation}
This threshold is determined by the classical bound $S_C$, the maximum quantum value $S_Q$, and the number of parties $n$. 
Here we only consider the positive-violation branch, $S'>S_C$. Accordingly, the correct-readout probability is restricted to the interval $p\in[\frac{1}{2},1]$ in the following analysis.
A Bell violation occurs for $p>p_{\rm cr}$, whereas for $p\leq p_{\rm cr}$ the observed Bell value does not exceed the corresponding classical bound. This conclusion applies not only to the Mermin inequality, but also to multipartite full-correlation inequalities with the same symmetric correlation-scaling structure. For the Mermin inequality, the classical bound $S_C$ exhibits a characteristic parity dependence. Substituting the corresponding expressions for $S_C$ and $S_Q$ from Table \ref{tab:table1} into Eq. (\ref{eq:23}), the critical correct-readout probability is
\begin{equation}
p_{\rm cr}^{Mer} =
\begin{cases}
    \frac{1 + 2^{(1-n)/{2n}}}{2}   & n \text{ odd} \\  
    \frac{1 + 2^{(2-n)/{2n}}}{2}   & n \text{ even}
\end{cases}.
\label{eq:24}
\end{equation}
A violation of the Mermin inequality is obtained when $p>p_{\rm cr}^{\rm Mer}$. As shown in Eq. (\ref{eq:24}), the even-$n$ threshold is higher than the odd-$n$ threshold, reflecting the suboptimality of the standard Mermin inequality for even numbers of parties.
To remove this even-$n$ suboptimality, we next consider the MABK inequality, which can be treated within the same correlation-scaling framework. Applying the MABK classical and quantum bounds to Eq. (\ref{eq:23}) gives

\begin{equation}
p_{\rm cr} ^{MK}=\frac{1+2^{ {(1-n)}/{2n }}}{2}.
\label{eq:25}
\end{equation}

Thus the MABK violation condition is $p>p_{\rm cr}^{\rm MK}$. For odd $n$, this threshold coincides with the Mermin threshold, whereas for even $n$ it provides the optimized multipartite Bell threshold.
For the Svetlichny inequality, the same substitution yields the critical correct-readout probability for genuine multipartite nonlocality certification,

\begin{equation}
p_{\rm cr}^{Sve}=\frac{1}{2} \left ( 1+  2^{-1/{2n} }\right ).
\label{eq:26}
\end{equation}
A Svetlichny violation is therefore observed only when $p>p_{\rm cr}^{\rm Sve}$.

\begin{figure}[t]
    \centering
    \includegraphics[width=\columnwidth]{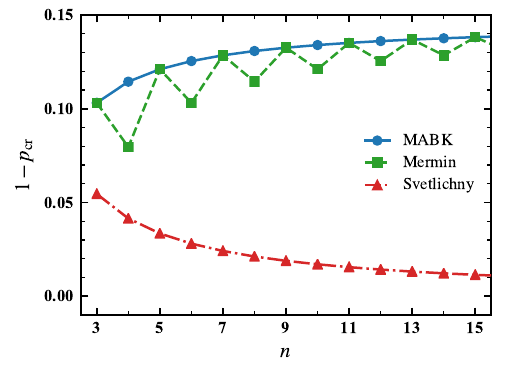} 
    \caption{Critical outcome-flipping probability $1-p_{\rm cr}$ required to violate different multipartite Bell inequalities as a function of the particle number $n$. The Mermin threshold (green squares) exhibits characteristic parity-dependent oscillations, reflecting its suboptimality for even $n$. In contrast, the MABK threshold (blue circles) increases monotonically and approaches the exact asymptotic limit of $(2-\sqrt{2})/4 \approx 0.1464$. The Svetlichny threshold (red triangles), used to certify genuine multipartite nonlocality (GMN), decreases monotonically with $n$.
    This indicates that the optimized MABK benchmark becomes increasingly tolerant
to stochastic outcome flipping, whereas certifying GMN becomes progressively
more demanding as the system size increases.  }
    \label{fig:Pcr}
\end{figure}

Fig.~\ref{fig:Pcr} illustrates that incoherent outcome flipping affects the Bell
violation through a scaling mechanism distinct from that of  coherent angular misalignment. In terms of the critical outcome-flipping probability $1-p_{\rm cr}$, the MABK threshold increases monotonically with $n$, asymptotically approaching $(2-\sqrt{2})/4$. This enhanced robustness arises because the exponentially growing multipartite Bell advantage partially counteracts the uniform correlation scaling $(2p-1)^n$. The Mermin threshold coincides with the MABK criterion for odd $n$, but suffers from parity-dependent oscillations for even $n$ due to its suboptimal classical bound. Conversely, the Svetlichny criterion becomes increasingly restrictive with
increasing $n$. Because the quantum-to-classical ratio of the Svetlichny operator remains fixed at $\sqrt{2}$,  certifying genuine multipartite nonlocality (GMN) becomes progressively more demanding as the system size scales up.

\subsection{Scaling and threshold structures under coherent and incoherent errors}

The preceding results show that coherent angular misalignment and incoherent outcome flipping affect multipartite Bell violations through different scaling mechanisms.
We compare these two measurement imperfections from
two complementary perspectives: their small-error scaling and the
resulting threshold-region structure.

First, the two error models exhibit different small-error scalings. For
coherent angular misalignment, local angular errors enter the
multipartite full-correlation terms through an accumulated phase. With
$m$ faulty measurement nodes, this phase is proportional to $m\theta$;
in the fully faulty case $m=n$, the critical misalignment angle therefore
scales as $\theta_{\rm cr}\sim O(1/n)$. This indicates that larger GHZ systems are increasingly sensitive to coherent systematic rotations.
For incoherent outcome flipping, the relevant effect is instead a stochastic attenuation of multipartite correlations. When all $n$ recorded outcomes are subject to the same independent readout error, the attenuation factor is $(2p-1)^n$. The corresponding threshold is determined by the competition between this attenuation and the quantum-to-classical ratio $S_Q/S_C$. For the MABK inequality and the odd-$n$ Mermin inequality, $S_Q/S_C=2^{(n-1)/2}$, which partly compensates the multiplicative decay of correlations. Consequently, the tolerable outcome-flipping probability $1-p_{\rm cr}$ approaches the finite large-$n$ limit $(2-\sqrt{2})/4\simeq 0.1464$.

Thus, coherent misalignment imposes an angular-precision requirement that becomes increasingly restrictive with increasing $n$, whereas incoherent outcome flipping retains a finite large-$n$ tolerance in terms of the critical outcome-flipping probability $1-p_{\rm cr}$ for the MABK and odd-$n$ Mermin criteria.

Second, the two models lead to different threshold-region structures. In
the fully faulty coherent case, the dependence on $n\theta$ makes the
critical angles periodic in the angular variable, so that violation and
non-violation regions appear alternately as $\theta$ varies over
$[0,\pi]$, as shown in Fig.~\ref{fig:2}. Over the full angular interval, the
individual coherent violation windows shrink as $1/n$, but their number
increases linearly with $n$, producing a narrower and denser reentrant
pattern. By contrast, incoherent outcome flipping uniformly rescales all
multipartite correlations by $(2p-1)^n$, which decreases monotonically
with $1-p$. Thus, for fixed $n$, the correct-readout probability defines
a single threshold separating the violation regime $p>p_{\rm cr}$ from
the non-violation regime $p\le p_{\rm cr}$.

\section{\label{paragraph4}Key-Rate Bounds under Measurement Imperfections}
The preceding section quantified how coherent angular misalignment and incoherent outcome flipping reduce the observable violations of multipartite Bell inequalities. These results provide thresholds for Bell-nonlocality certification. However, Bell-nonlocality certification alone does not guarantee that a positive Devetak--Winter lower bound can be obtained. Key-rate estimation imposes additional requirements: the observed violation must provide a bound on Eve's information, and the key-generation measurements must maintain a sufficiently low quantum bit error rate (QBER) among the legitimate parties. We therefore relate the degraded Bell values derived above to asymptotic Devetak--Winter key-rate bounds within a convex-combination attack model \cite{Devetak2005207}.

In the broader context of multipartite device-independent cryptography, recent studies have addressed long-distance conference key agreement and quantum key activation \cite{Ishihara2025,Ulu2025}. We focus here on a multipartite secret-sharing scenario in which the parties share an $n$-partite state and generate a shared key from full-correlation measurements.
The Bell-test rounds are used to estimate the observed
Svetlichny violation and thereby certify genuine
multipartite nonlocality, whereas the key-generation
rounds rely on deterministic GHZ correlations.

We model Eve's eavesdropping strategy as a convex-combination attack, assuming that she exercises control over the entanglement source. Upon the public announcement of the measurement settings by the legitimate parties in each round, Eve is assumed to have full access to information from the local measurement stations. Under this attack model, the observed degraded Bell value can be written as a convex combination of the local and nonlocal contributions,

\begin{equation}
    S'=q_L S_C+q_{\rm NL}S_Q ,
\label{eq:27}
\end{equation}
where $q_L\in[0,1]$ denotes the weight of the threshold component accessible to Eve, and $q_{\rm NL}=1-q_L$ denotes the weight of the maximally violating component. Under this model, Eve is assumed to have full information about the threshold component, while her optimal guess for the maximally violating component is random. Therefore, the local weight $q_L$ provides an analytical link between the degraded Bell value and Eve's information. Solving Eq. (\ref{eq:27}) for $q_L$, we obtain

\begin{equation}
\begin{split}
    q_L=\frac{S_Q -S'}{{S_Q} -S_C}=\frac{2-2\chi_1}{2-\sqrt{2}},
\end{split}
\label{eq:28}
\end{equation}
where $\chi_1=S'/S_Q$ reflects the relationship between imperfect measurement parameters and the quantum violation value. 
We use the Devetak--Winter formula \cite{Devetak2005207} to estimate the asymptotic extractable key rate
\begin{equation}
    r_{DW} \geq H\left(A_1 | E\right) - H\left(A_1 | A_2, A_3 \dots A_n\right).
    \label{eq:29}
\end{equation}

The quantity $H (A_1 | E) $ characterizes Eve's uncertainty about the
outcome of party $A_1 $, whereas $H(A_1|A_2,A_3 \dots A_n)$ characterizes the residual uncertainty of $A_1 $ conditioned on the outcomes of the other legitimate parties. Under the convex-combination attack model, Eve possesses full information about the local component, whereas her optimal guessing probability for the nonlocal component is $\frac{1}{2}$ \cite{Xiang2023S,Xiang2023}. The corresponding conditional entropies are therefore given by 

\begin{subequations}
\label{eq:30}
\begin{align}
&\begin{aligned}
    H(A_1|E) &= - \sum_{(A_1,E)} P(E) P(A_1|E) \log_2 P(A_1|E)   \\
    &=h \left( \frac{1 + q_L}{2} \right),
\end{aligned}
\label{eq:30a}\\
    & H(A_1 | A_2, A_3 \dots A_n)=h \left( Q \right).
\label{eq:30b}
\end{align}
\end{subequations}
Here, $h(x) = -x \log_2 x - (1-x) \log_2 (1-x)$ represents the binary entropy, and $Q$ denotes the quantum bit error rate (QBER). When generating a key, the legitimate parties choose the agreed measurement settings, perform joint measurements, and generate a shared key.
For a valid key-generation setting, the ideal GHZ full correlation is deterministic. In the key-generation process, any stabilizer of the GHZ state can be selected to generate the key. Consequently, for any joint measurement, the QBER can always be written in the following form: \cite{PhysRevA.82.012304,Epping_2017}

\begin{equation}
    Q=\frac{1-\chi_2}{2},
    \label{eq:31}
\end{equation}
where $\chi_2 = |\left \langle A_1A_2 \dots A_n \right \rangle|$ characterizes the degradation factor associated with the QBER among the legitimate parties. Under ideal measurements, the stabilizers of the $n$-GHZ state yield deterministic eigenvalues of $\pm 1$, so the corresponding QBER vanishes. In the presence of imperfect measurements, however, the relevant expectation values deviate from their ideal deterministic values, thereby producing a nonzero error rate. The asymptotic secret key rate formula becomes

\begin{equation}
    r_{DW}\geq h\left(\frac{1}{2}+\frac{1-\chi_1}{2-\sqrt{2}}\right)-h\left(\frac{1-\chi_2}{2}\right).
    \label{eq:32}
\end{equation}

\subsection{Influence of coherent angular misalignment on secret key rates }

In the quantum secret-sharing (QSS) setting considered here, measurement-angle deviations directly affect the QBER. To extract a deterministic result from an $n$-partite GHZ state satisfying $(A_1 A_2 \dots A_n)|\text{GHZ}\rangle =\pm |\text{GHZ}\rangle)$, the legitimate parties perform joint measurements in the $x-y$ plane. In this setting, an odd number of $\sigma_y$ operators must be applied when generating keys.

By specifying these measurement bases, the associated QBER can be evaluated, which is essential for deriving the final secret key rate. In the tripartite scenario ($n=3$), only four valid measurement configurations satisfy this condition: $\sigma_y\sigma_x\sigma_x$, $\sigma_x\sigma_y\sigma_x$, $\sigma_x\sigma_x\sigma_y$, and $\sigma_y\sigma_y\sigma_y$. 
Under the common-rotation convention introduced in Sec.~\ref{3.A}, the expressions for the resulting QBER are detailed in Table \ref{tab:table2}.

\begin{table}[htp]
\centering
\caption{\label{tab:table2}
Orthogonal measurement-angle settings and QBER for the three-partite case $n=3$.}
\renewcommand{\arraystretch}{1.15}
\begin{tabular*}{\columnwidth}{@{}c@{\extracolsep{\fill}}cccc@{}}
\toprule
Setting & $A_1$ & $A_2$ & $A_3$ & $Q$ \\
\midrule
$\sigma_y\sigma_x\sigma_x$
& $\frac{\pi}{2}+\theta$
& $\theta$
& $\theta$
& $\frac{1-\cos(3\theta)}{2}$ \\

$\sigma_x\sigma_y\sigma_x$
& $\theta$
& $\frac{\pi}{2}+\theta$
& $\theta$
& $\frac{1-\cos(3\theta)}{2}$ \\

$\sigma_x\sigma_x\sigma_y$
& $\theta$
& $\theta$
& $\frac{\pi}{2}+\theta$
& $\frac{1-\cos(3\theta)}{2}$ \\

$\sigma_y\sigma_y\sigma_y$
& $\frac{\pi}{2}+\theta$
& $\frac{\pi}{2}+\theta$
& $\frac{\pi}{2}+\theta$
& $\frac{1-\cos(3\theta)}{2}$ \\
\bottomrule
\end{tabular*}
\end{table}

Under the condition of measurement imperfections where both measurement directions experience an identical deflection, the same error rate is ensured regardless of the choice of measurement bases. Specifically, as illustrated in the tripartite scenario ($n=3$), selecting different measurement configurations (e.g., $\sigma_x\sigma_x\sigma_y$ and $\sigma_y\sigma_y\sigma_y$) yields the same QBER.

It is worth noting that $\chi_1$ is associated with all correlation terms appearing in the Bell inequality and therefore characterizes the overall degradation of the Bell violation. In contrast, $\chi_2$ is determined only by the particular stabilizer correlation chosen for key generation and is used to quantify the QBER under that key-generation setting. Thus, $\chi_1$ and $\chi_2$ do not have exactly the same physical meaning: the former describes the global correlation degradation relevant to nonlocality certification, whereas the latter describes the error behavior associated with a specific key-generation basis. When the two measurement directions undergo identical angular deflection, the QBER-related degradation factor becomes $\chi_2=\cos(n\theta)$. Since this factor is identical to the Bell-violation degradation factor, we have $\chi_1=\chi_2=\chi$. Substituting this relation into the Devetak--Winter key-rate expression gives the condition for positive key generation as $\chi>\frac{2}{4-\sqrt{2}}$. Therefore, the threshold measurement parameter $\theta_{\rm th}$ is

\begin{equation}
    \theta_{\rm th}=\frac{\pm \arccos(\frac{2}{4-\sqrt{2}})+2k\pi}{n}.
    \label{eq:33}
\end{equation}

For coherent angular misalignment, positive key generation is allowed only within periodic angular windows centered at $\theta_k=2k\pi/n$. The half-width of each key-generation window is determined by $\arccos[2/(4-\sqrt{2})]/n \approx 0.218 \pi/n$, which is smaller than the half-width $0.25 \pi/n$ of the corresponding Bell-violation window for genuine multipartite nonlocality. Therefore, secure key generation occupies a narrower parameter region than inequality violation and requires higher angular measurement accuracy. Furthermore, the secret key rate reaches its maximum value of unity at the center of each window, where the inequality violation is also maximized.

\subsection{Influence of incoherent outcome flipping on secret key rates }

For incoherent outcome flipping, the measurement basis is correctly implemented, but each recorded binary outcome is flipped with probability $1-p$, where $p$ is the correct-readout probability. Because this stochastic error is independent of the measurement setting, all valid key-generation correlations acquire the same scaling factor $(2p-1)^n$. Thus, for any key-generation setting that ideally gives a deterministic GHZ full correlation, the QBER is $Q=\frac{1-(2p-1)^n}{2} = \frac{1-\chi_2}{2}$ with $\chi_2=(2p-1)^n$.

The relationship between the secret key rate  and the measurement imperfection is again given by Eq. (\ref{eq:32}), with $\chi_1=\chi_2$. 
Setting the key rate
to zero yields an analytical threshold for the correct-readout probability,
\begin{equation}
    p_{\rm th} =\frac{1}{2} \times \left [ \left ( \frac{2}{4-\sqrt{2}}  \right ) ^{\frac{1}{n} } + 1 \right ].
\label{eq:34}
\end{equation}

For probabilistic outcome flipping, the condition for positive key generation is characterized by a critical value of the correct-readout probability $p$, rather than by periodic angular windows. The threshold $p_{\rm th}$ required for secure key generation, given in Eq. (\ref{eq:34}), is more stringent than the threshold $p_{\rm cr}$ required for genuine multipartite nonlocality certification, given in Eq. (\ref{eq:26}). Since $p$ represents the probability of obtaining the correct outcome, this comparison shows that secure key generation requires a higher correct-outcome probability than genuine multipartite nonlocality certification. Consequently, there exists a parameter region in which genuine multipartite nonlocality can still be certified, while a positive secret key rate cannot be obtained.

\subsection{Werner states with imperfect measurements}

The preceding analysis assumes that the shared state is an ideal GHZ state and that the degradation arises solely from measurement imperfections. In realistic implementations, however, source and channel noise also reduce the visibility of multipartite correlations.
To incorporate these effects, we consider an $n$-partite Werner state, which provides a standard visibility-noise model for studying the relation between mixed-state entanglement and Bell nonlocality \cite{Werner1989,Luo2018}
\begin{equation}
\rho^{v}  = {v} |\psi \rangle \langle\psi| + (1 - {v}) \frac{\mathbb{I}}{2^{n}}.
\label{eq:35}
\end{equation}
Here, $v\in \left [ 0,1\right]$ is the visibility of the GHZ-state component. The visibility $v$ rescales the ideal GHZ full correlations, since the maximally mixed component does not contribute to the Bell correlators or to the deterministic key-generation correlations. The legitimate parties share the Werner state under imperfect measurements, with only the GHZ component contributing to the relevant full correlations. Let $S'$ denote the Bell value degraded only by measurement imperfections.
For the Werner state, the observed Bell value becomes $S_v'=vS'$. The corresponding local weight in the convex-combination model is
\begin{equation}
q_L^{v}=\frac{S_Q-vS'}{S_Q-S_C}.
\label{eq:36}
\end{equation}

The QBER is similarly rescaled by the Werner-state visibility and is given by
\begin{equation}
Q^{v}=\frac{1-v\chi}{2}.
\label{eq:37}
\end{equation}

Substituting Eqs.~(\ref{eq:36}) and (\ref{eq:37}) into the Devetak--Winter bound gives the Werner-state key-rate expression
\begin{equation}
r_{\rm DW}\ge
h\left(\frac{1}{2}+\frac{1-v\chi}{2-\sqrt{2}}\right)
-
h\left(\frac{1-v\chi}{2}\right),
\label{eq:38}
\end{equation}
where $\chi$ denotes the correlation reduction factor induced by the
measurement imperfection. For incoherent outcome flipping,
$\chi=(2p-1)^n$, whereas for coherent angular misalignment with $m$
faulty nodes in the common-rotation case, $\chi=\cos(m\theta)$.

The corresponding visibility thresholds for a positive Devetak--Winter lower bound are therefore given by
\begin{subequations}
\label{eq:39}
\begin{align}
v_{\rm th}^{p}&=\frac{2}{(4-\sqrt{2})(2p-1)^n},
\label{eq:39a}
\\[2mm]
v_{\rm th}^{\theta}&=\frac{2}{(4-\sqrt{2})\cos(n\theta)}.
\label{eq:39b}
\end{align}
\end{subequations}
Here, Eq. (\ref{eq:39a}) corresponds to incoherent outcome flipping, whereas Eq. (\ref{eq:39b}) corresponds to coherent angular misalignment. These thresholds correspond to physical Werner-state visibilities only when $0\leq v_{\rm th}\leq1$.

\begin{figure}[t]
    \centering
    \includegraphics[width=\columnwidth]{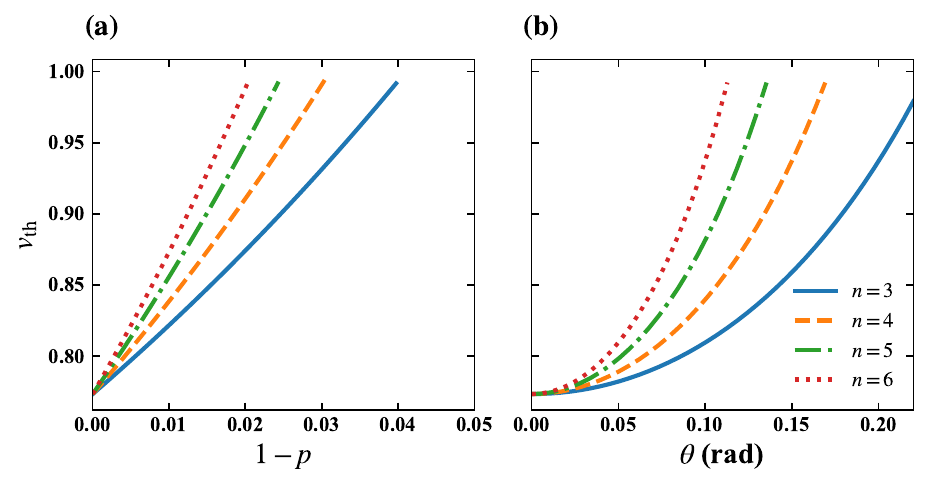} 
    \caption{ Threshold visibility $v_{\rm th}$ required for positive secret key generation in $n$-partite Werner states under imperfect measurements. The region above each curve corresponds to $v>v_{\rm th}$. (a) $v_{\rm th}^{p}$ as a function of the outcome-flipping probability $1-p$. (b) $v_{\rm th}^{\theta}$ as a function of the coherent misalignment angle $\theta$. In both cases, larger $n$ requires higher visibility to tolerate the same level of measurement imperfection.
    }
    \label{fig:5}
\end{figure}

Fig.~\ref{fig:5} shows the resulting visibility thresholds as functions of the measurement-imperfection parameters for $n=3,4,5,6$. The secure region lies above each curve. In Fig.~\ref{fig:5} (a), $v_{\rm th}^{p}$ increases with the outcome-flipping probability $1-p$, reflecting the stochastic attenuation factor $(2p-1)^n$. In Fig.~\ref{fig:5} (b), $v_{\rm th}^{\theta}$ increases with the coherent misalignment angle $\theta$, reflecting the accumulated phase factor $\cos(n\theta)$. In both cases, increasing the number of parties raises the visibility required to tolerate the same level of measurement imperfection.

\section{\label{paragraph5}Conclusion and Discussion}

In this work, we have studied the effects of coherent angular misalignment and incoherent outcome flipping on multipartite Bell nonlocality for the Mermin, Svetlichny, and MABK inequalities in $n$-partite GHZ Bell-test scenarios. 
For coherent angular misalignment, under the orthogonality-preserving condition and in the fully faulty case $m=n$, the corresponding critical angles are obtained from the condition $\theta_{\rm cr} = (\pm \arccos(S_C/S_Q) + 2k\pi)/n$.
These critical angles define periodic violation windows centered at $\theta_k=2k\pi/n$. In the large-$n$ limit, the half-widths of these windows approach $\Delta\theta_{\rm Mer/MK}\simeq \pi/(2n)$ for the Mermin and MABK inequalities and $\Delta\theta_{\rm Sve}\simeq \pi/(4n)$ for the Svetlichny inequality, giving $\Delta\theta_{\rm Mer/MK}\simeq 2\Delta\theta_{\rm Sve}$. Therefore, within the considered angular interval [0,$\pi$], the total violation length approaches $\pi/2$ for the Mermin and MABK inequalities, whereas it approaches $\pi/4$ for the Svetlichny inequality.
By contrast, for incoherent outcome flipping, the corresponding critical threshold follows from the condition $p_{\rm cr} = ((S_C/S_Q)^{\frac{1}{n}}+1)/2$.
Since the degraded Bell value varies monotonically with the correct-readout probability $p$, the outcome-flipping model provides a single threshold criterion separating the violation and non-violation regimes. For the MABK inequality and each fixed-parity branch of the Mermin inequality, the tolerable outcome-flipping probability $1-p_{\rm cr}$ increases with $n$ and approaches $(2-\sqrt{2})/4$. 
In contrast, for the Svetlichny inequality, $1-p_{\rm cr}$ decreases with $n$ and approaches zero as $n\to\infty$.

We further connect the degraded Svetlichny values with the asymptotic Devetak--Winter key-rate bound under the convex-combination attack model.
For coherent angular misalignment, under the orthogonality-preserving condition and in the fully faulty case $m=n$, the threshold angle for positive key generation follows from the condition $\theta_{\rm th} = (\pm \arccos[2/(4-\sqrt{2})] + 2k\pi)/n$.
In this case, positive key generation is allowed only within periodic angular windows centered at $\theta_k=2k\pi/n$, whose half-width is determined by $\arccos[2/(4-\sqrt{2})]/n$. 
For incoherent outcome flipping, the asymptotic Devetak--Winter key-rate bound gives a critical readout threshold following from the condition $p_{\rm th}=([2/(4-\sqrt{2})]^{\frac{1}{n}}+1)/2$.
Since the key-rate condition varies monotonically with the correct-readout probability $p$, this threshold provides a clear criterion separating the positive-key and no-key regimes. 
In both coherent angular misalignment and incoherent outcome flipping, the generation of a positive key requires stricter tolerance to measurement imperfections than the threshold for genuine multipartite nonlocality violation.
By considering key generation with Werner states, we obtain the relationship between the key rate, state visibility, and measurement imperfections, showing that larger $n$ requires higher visibility to tolerate the same measurement error.

The present analysis can be extended to non-isotropic noise models and to more complex entanglement structures beyond GHZ states, such as graph states and high-dimensional qudit systems. It would also be useful to combine the present analytical models with experimental calibration and mitigation strategies: coherent angular misalignment calls for basis calibration or rotation compensation, whereas incoherent outcome flipping can be addressed by readout-error mitigation. Future work should also investigate multipartite measurement imperfections under more general eavesdropping strategies beyond the convex-combination attack model.

\section*{Acknowledgments}
This work was supported by the National Natural Science Foundation of China (Grant Nos. 12464064 and 12165020), the Major Science and Technology Project of Yunnan Province (Grant No. 202502AD080015), and the Yunnan Fundamental Research Project (Grant No. 202401AT070448). L.-J. Wang acknowledges support from the Xing Dian Talent Program (Grant No. C619300A090). W.-L. Qiao appreciates support from the Postgraduate Research Innovation Fund Project of Yunnan University (Grant No. KC-252511703).

\section*{Data availability}
No experimental datasets were generated in this work. The numerical values used to produce the figures are obtained from the analytical expressions presented in the text.

\bibliographystyle{quantum}

\bibliography{apssamp}

\onecolumn
\appendix

\section{\label{Appendix:A}Different parties experience distinct measurement-direction misalignment}

Here, we consider party-dependent coherent misalignment, where different parties experience distinct systematic angular deviations. We assume that the intended measurement settings are correctly assigned, but the measurement instruments have fixed calibration offsets. Consequently, the deflection error for a given party remains constant across different measurement runs, while these systematic errors may vary among different parties. 

To analyze this quantitatively, let $\phi_j$ denote the ideal measurement angle in the $x-y$ plane for the $j$-th party, and let $\delta_j$ represent the corresponding calibration offset. The actual measurement direction for that party is thus modified to $\theta_j=\phi_j+\delta_j$, with the corresponding measurement observable parameterized as
\begin{equation}
M(\theta_j)=\cos\theta_j\sigma_x+\sin\theta_j\sigma_y.
\label{eq:A1}
\end{equation}
For any $n$ parties sharing an $n$-qubit GHZ state, the expectation value of their joint measurement is highly dependent on the phase structure of the initial state. 
When the system is prepared in the standard GHZ state $\frac{1}{\sqrt{2}}(|0\rangle^{\otimes n} + |1\rangle^{\otimes n})$, the joint expectation value can be analytically expressed as the cosine of the sum of all individual measurement angles. 
By substituting the actual measurement angles with the intended angles and their respective systematic errors, the cumulative effect can be explicitly formulated as
\begin{equation}
E=\cos\left(\sum_{j=1}^n\theta_j\right)=\cos\left[\sum_{j=1}^n(\phi_j+\delta_j)\right].
\label{eq:A2}
\end{equation}
Conversely, when the system is in the GHZ state with a complex phase, $\frac{1}{\sqrt{2}}(|0\rangle^{\otimes n} + i|1\rangle^{\otimes n})$, the presence of the imaginary unit $i$ in the interference terms transforms the joint expectation value into the corresponding sine function. Under the influence of the same systematic errors, this shifted expectation value is given by

\begin{equation}
E = \sin\left(\sum_{j=1}^n \theta_j\right) = \sin\left[\sum_{j=1}^n (\phi_j + \delta_j)\right].
\label{eq:A3}
\end{equation}

For a single full-correlation stabilizer term, the systematic angular
errors therefore enter through the accumulated phase. If the ideal
correlation reaches its extremal value, the corresponding degraded
correlation can be written as
\begin{equation}
    E'=E_{\rm id}\cos\left(\sum_{j=1}^n\delta_j\right),
    \label{eq:A4}
\end{equation}
where \(E_{\rm id}\) denotes the ideal full-correlation value.

This expression illustrates the phase-accumulation mechanism for
party-dependent coherent misalignment. For a full Bell operator, the
degraded Bell value is obtained by summing the corresponding shifted
full-correlation terms with the coefficients of that particular
inequality.

\section{\label{Appendix:B}General symmetric multipartite inequalities}

For the class of two-input, two-output full-correlation Bell inequalities, the expectation values of the operators can be expressed in terms of probabilities. Taking a correlation term in the CHSH inequality as an example, the term $\left \langle A_0B_0 \right \rangle$ can be rewritten as $\left \langle A_0B_0 \right \rangle =P(0,0|0,0)+P(1,1|0,0)-P(0,1|0,0)-P(1,0|0,0)=2P(a\oplus b=0|0,0)-1$. Here, $a$ and $b$ represent the measurement outcomes of the parties. By extension, various Bell inequalities can be reformulated as linear combinations of joint probabilities. In an $n$-partite problem, when the measurement outcomes satisfy specific logical relations, the terms with positive coefficients and those with negative coefficients can be written respectively as

\begin{equation}
1 - 2 \sum_{\{X_i\} \in \mathcal{S}} P(X_1\oplus X_2\oplus  \dots  \oplus X_n=\eta|x_1,x_2 \dots x_n) ,
 \label{eq:B1}
\end{equation}

\begin{equation}
2 \sum_{\{X_i\} \in \mathcal{S}} P(X_1\oplus X_2\oplus  \dots  \oplus X_n=\eta|x_1,x_2 \dots x_n) -1.
 \label{eq:B2}
\end{equation}

Let $X_1,X_2 \dots X_n$ be the measurement outcomes of the parties, and let $\mathcal{S}$ denote the set of solutions that satisfy the predefined logical conditions. The variables $x_1,x_2 \dots x_n$ represent the measurement settings of different parties, corresponding to their respective measurement directions. The symbol $\eta$ characterizes the relation between the measurement outcomes. Consequently, the corresponding full-correlation contribution can be expressed as

\begin{equation}
    2\sum_{ \left \{  x_i\right \} } P(X_1\oplus X_2\oplus  \dots \oplus  X_n=\eta|x_1,x_2 \dots x_n)-C.
    \label{eq:B3}
\end{equation}

Here, $C$ is a constant associated with the number of terms in the inequality. We define Bell inequalities that can be expressed in such a probabilistic form, and whose operator coefficients have equal absolute values, as Bell inequalities with a symmetric structure. In the main text, we derive the threshold measurement parameter for both Bell-inequality violation and key generation under imperfect measurements for the standard multipartite Mermin and MABK inequalities. The results indicate that this threshold measurement parameter is determined by the quantum upper bound and the classical bound restricted by LHV models. This conclusion applies not only to Mermin and MABK inequalities but to all inequalities with a symmetric structure; however, it cannot be extended to those with non-symmetric structures. While well-known inequalities such as CHSH, Svetlichny, Mermin, and MABK all possess symmetric structures, Hardy's inequality for three or more parties does not belong to this category.

\end{document}